\documentstyle[12pt]{article}

\newcommand{\eq}[1]{(\ref{#1})}
\newcommand{\diff}{\partial}
\newcommand{\beq}{\begin{equation}}
\newcommand{\eeq}{\end{equation}}
\newcommand{\beqn}{\begin{eqnarray}}
\newcommand{\eeqn}{\end{eqnarray}}

\def\NP{ Nucl.~Phys.}
\def\PL{ Phys.~Lett.}
\def\PRL{ Phys.~Rev.~Lett.}
\def\PRp{ Phys.~Rep.}
\def\PR{ Phys.~Rev.}

\title{
\vspace{-1.5cm}
\begin{flushright}
\begin{Large}
ITEP-95-34\\
hep-th/9506026
\end{Large}
\end{flushright}
\vspace{1.5cm}
Instantons and Monopoles in Maximal Abelian Projection of $SU(2)$
Gluodynamics}

\author{
{\sc M.N.~Chernodub\thanks{E--mail address:
{\sc chernodub@vxitep.itep.ru}
}
and F.V.~Gubarev
\vspace{0.5cm}
}\\
{\sl ITEP, Moscow, 117259, Russia}\\
{\sl and}\\
{\sl MIPT, Dolgoprudny, Moscow region, Russia}
}

\date{}

\begin{document}
\bibliographystyle{bibstand}
\maketitle
\thispagestyle{empty}

\begin{abstract}
We show that the instantons induce the abelian monopoles in the Maximal
Abelian Projection of $SU(2)$ gluodynamics. As an example we consider the
case of one instanton and the case of a set of instantons arranged along a
straight line.  The abelian monopoles which are induced by instantons may
play some role in the confinement scenario.
\end{abstract}


\newpage

\section{Introduction}

 The confinement phenomena in QCD is one of the most important
problems in the nonabelian theories.
Various instanton models of QCD yield
vanishing string tension for the QCD string \cite{Inst-String}, however, the
instantons
seem to play an important role. For
instance the correlation functions of various colourless quark bound states
are well described by the instanton contributions \cite{ChGrHuNe94}.

The explanation of the confinement phenomena was proposed by
't~Hooft \cite{tHo76} and Mandelstam \cite{Man76}, who conjecture
that the infrared properties of confining QCD vacuum are similar to those of
the (dual) superconductor. The most convenient way to think about this
analogy is to partially fix the $SU(N)$ gauge degrees of freedom leaving
${[U(1)]}^{N-1}$ group unfixed \cite{tHo81}. Such partial gauge is usually
called abelian projection. Under the abelian transformations, the
diagonal elements of the gluon field transform as gauge fields and due to
the compactness of the $U(1)$ gauge group, the abelian monopoles exist. If
they are condensed, the string between the coloured charges is formed as the
dual analogue of the Abrikosov string in a superconductor, the monopoles
playing the role of the Cooper pairs \cite{tHo76,Man76}.

Many numerical experiments demonstrate the dual superconductor mechanism in
$SU(2)$ lattice gauge theory (for a review see \cite{Suz93}); the most part
of the results was obtained in so called Maximal Abelian (MA) projection
\cite{KrLaScWi87}. However, there exists the Abelian projection in which the
role of monopoles is played by the other topological objects - by the
``minopoles'' \cite{ChPoVe95}.

Below we show that
several instantons in the 't Hooft anzatz, with the centers placed on the
straight line, lead to the
abelian monopole current along this line.
For the sake of simplicity we discuss only the
$SU(2)$ gauge group.

\section{From Instanton to Abelian Monopole}

The $SU(2)$ instanton field configuration in the 't~Hooft
anzatz is given by the equation

\beq
        A^a_\mu = \frac{1}{g} \bar{\eta}^a_{\mu\nu} \diff_\nu f(x)\,,
        \label{tHooft:anz}
\eeq
where $\bar{\eta}^a_{\mu\nu}$ is 't~Hooft symbol and $g$ is coupling
constant. The function $f(x)$ is given by

\beq
  f(x) = f^I(x) =
  \ln \left[1+\frac{\rho^2}{t^2 + r^2} \right]\,,
  \label{inst}
\eeq
where $\rho$ is the size of the instanton, $t$ -- time, $r$ -- spatial
radius.

The gauge conditions which define the MA projection are given by the
formula \cite{KrLaScWi87}:

\beq
   (\diff_\mu \pm i g A^3_\mu) A^\pm_\mu = 0\,, \label{MAA:cond}
\eeq
where $A^\pm_\mu = A^1_\mu \pm i A^2_\mu$, the $SU(2)$ generators are
$t^a = \sigma^a \slash 2$ and $\sigma^a$ are the Pauli matrices.
Let us rotate the instanton field (\ref{tHooft:anz},\ref{inst}) by the
$SU(2)$ matrix $\Omega$,

\beqn
 \Omega = { \cos\phi\, e^{i \theta}   \quad   \sin\phi\, e^{i \chi} \choose
          - \sin\phi\, e^{- i \chi}   \quad   \cos\phi\, e^{- i \theta} }\,,
 \label{Omega}
\eeqn
where

\beq
   \chi = \Delta - \alpha \slash 2\,, \quad
   \theta = - \Delta - \alpha \slash 2\,, \quad
   \phi = \gamma \slash 2\,. \label{angles}
\eeq
Here $\alpha$ and $\gamma$ are azimuthal and polar angles of the reference
system in the given time--slice, $\Delta(x)$ is an arbitrary function. As
it can be easily checked, the instanton field $A$ which is rotated by this
matrix $\Omega$

\beq
  A_\mu \to A^{(\Omega)}_\mu = \Omega^+ A_\mu \Omega - \frac{i}{g} \,
  \Omega^+ \diff_\mu \Omega\,,
  \label{GaugeTransf}
\eeq
satisfies the MA projection conditions \eq{MAA:cond}.

The field strength tensor

\beq
  G_{\mu\nu}[A] = \diff_\mu A_\nu - \diff_\nu A_\mu + i g \, [A_\mu,A_\nu]
  \label{Gmunu}
\eeq
with the transformed field $A^{(\Omega)}_\mu$ satisfies the equation

\beq
   \epsilon_{\mu\nu\alpha\beta} \diff_{\nu} G_{\alpha\beta}[A^{(\Omega)}] =
   \frac{2 \pi}{g} j_{\mu} \cdot \sigma^3\,, \label{j}
\eeq
where $j_{\mu}$ is given by the following equation:

\beq
   j_{\mu}(x) = \delta^{(3)}({\bf r}) \cdot \delta_{\mu 0}\,.
   \label{mon:current}
\eeq
The equation \eq{j} is nothing but the definition of the monopole
singularities in the gauge field $A^3_\mu$, since the only third component
$G^3[A^{(\Omega)}]$ contributes\footnote{Note, that the field
strength tensor \eq{Gmunu} transforms under singular gauge transformations
$\Omega$ as follows: $G_{\mu\nu} \to G_{\mu\nu}[A^{(\Omega)}] = \Omega^+
G_{\mu\nu}[A] \Omega + G^{sing}_{\mu\nu}[\Omega]$, where
$G^{sing}_{\mu\nu}[\Omega] = - i \Omega^+(x) [ \diff_{\mu} \diff_{\nu} -
\diff_{\nu} \diff_{\mu} ] \Omega(x)$. The matrix \eq{Omega} is
singular and therefore $G^{sing}_{\mu\nu}[\Omega]$ is not zero. It can
be also shown that only $G^{sing}_{\mu\nu}[\Omega]$ contributes
to the {\it r.h.s.} of eq.\eq{j}.
}
to the {\it r.h.s} of
eq.\eq{j}. The monopole current
$j_\mu(x)$ in eq.\eq{mon:current} is the straight line along the
time--direction which crosses the center of the instanton (which is the
center of our reference system).

Note, that the monopole trajectory does not depend on the choice of the
function $\Delta(x)$, which was used in the definition of the matrix
$\Omega$, eqs.\eq{Omega} and \eq{angles}. As it can be shown, this function
corresponds to the residual $U(1)$ degrees of freedom we leave
unfixed. $\Delta(x)$ does not affect $U(1)$--invariant
quantities while $U(1)$--variant quantities ({\it e.g.}, the
position of the Dirac string which is attached to the abelian monopole)
changes if the function $\Delta(x)$ is changed.

The choice \eq{angles} for the $SU(2)$--angles $\theta$, $\chi$ and $\phi$
is not unique. Let us choose an arbitrary unit vector $n_\mu$ and define
three vectors $m^{(a)}_\mu$, $a=1,2,3$ as follows:

\beq
   m^{(k)}_\mu = \bar{\eta}^a_{\mu\nu} n_\nu\,.
   \label{m}
\eeq
Using the properties of the 't~Hooft symbols $\bar{\eta}^a_{\mu\nu}$ we can
easily derive that vectors $n$ and $m^{(a)}$ form the basis in $4D$
Euclidean space--time:

\beq
  m^{(k)}_\mu m^{(l)}_\mu = \delta^{kl}\,;\quad
  m^{(k)}_\mu n_\mu = 0\,\quad k=1,2,3\,.
  \label{3d}
\eeq

Let us choose the $SU(2)$--angles $\theta$, $\chi$ and $\phi$ as in
eq.\eq{angles} but with the azimuthal, $\alpha$, and polar, $\gamma$, angles
are defined with respect to the new $3D$ basis $\{ m^{(a)}_\mu \}$. We call
the matrix $\Omega$, eq.\eq{Omega}, with new angles $\theta$, $\chi$ and
$\phi$ as $\Omega[n]$.  It can be easily verified that the transformation of
the instanton field (\ref{tHooft:anz},\ref{inst}) by the matrix $\Omega[n]$
satisfies the MA projection conditions \eq{MAA:cond}.

Repeating all the above calculations, but with new matrix $\Omega$ we get
for the abelian monopole current which has to be defined from eq.\eq{j}:

\beq
   j_{\mu} = \delta^{(3)}({\bf y}) \cdot n_\mu\,,
   \label{mon:current2}
\eeq
where $y^k = m^k_\nu x_\nu$. The monopole trajectory is the straight line
which crosses the center of the instanton and has the direction $n_\mu$.

\section{Two and Many Instantons}

The $N$--instanton field configuration $A^{N \cdot I}$ may be written in the
't~Hooft anzatz \eq{tHooft:anz}, where the function $f(x)$ is given by the
formula:

\beq
  f(x) = f^{N \cdot I}(x) = \ln \left[1+ \sum^N_{i = 1} \frac{\rho^2_i}{{(x -
  x_i)}^2} \right]\,.
  \label{Ninst}
\eeq
Here $x_i$ is the position of the center of $i$--th instanton, all the
instantons have the same colour orientations.

Consider first the case of two instantons, $N = 2$. Let us rotate the field
(\ref{tHooft:anz},\ref{Ninst}) by the matrix $\Omega[n]$, where the vector
$n_\mu$ is oriented along the line which connects the centers $x_1$ and
$x_2$ of these instantons. The direct evaluation shows that the rotated
field satisfies the MA projection conditions \eq{MAA:cond}. Substituting
the field $A = A^{N \cdot I}$ and the matrix $\Omega = \Omega[n]$ into
eq.\eq{j} we finally get for the abelian monopole current the expression
\eq{mon:current2}. Note, that the sizes $\rho_i$ of the considered
instantons are not important.

Thus we can draw the conclusion: any two instantons in the 't Hooft anzatz
bear in the MA projection the straight--line abelian monopole
current $j_\mu$ which goes through the centers of these instantons. Note
that we did not prove the uniqueness of the solution of the equation
(\ref{MAA:cond}) with respect to the gauge transformations \eq{GaugeTransf},
the existence of the other monopoles trajectories are not excluded.

The already described matrix $\Omega[n]$ rotates an arbitrary configuration
of instantons within the MA projection, provided that the centers of all
these instantons are on the same straight line and the colour orientations
are also the same. As in the two--instanton case, the sizes of instantons
are irrelevant and the resulting  abelian monopole trajectory goes through
the centers of instantons.

The particular case of the considered instanton configuration
corresponds to the BPS--monopole \cite{BPS}: the distance between any two
neighbour centres should be fixed and the sizes of the all instantons should
be equal to infinity. The BPS--monopole
trajectory (which is the straight line) coincides with the trajectory
of the abelian monopole.
This case was already discussed in ref. \cite{SmSi91}.

There are more general field configurations which give in the MA
projection, the abelian monopole trajectories which are arbitrary straight
lines with the direction $n_\mu$. These field configurations may be
represented in the 't~Hooft anzatz \eq{tHooft:anz} and the function $F(x)$
being dependent only on the combinations $t' = n \cdot x$ and $r' =
\sqrt{x^2 - {(n \cdot x)}^2}$. The resulting monopole lines cross the center
of our reference system.

\section{Conclusions}

Our calculations show that the single instanton field configuration can be
rotated within the MA projection by various gauge transformations each of
which is parametrized by arbitrary vector~$n_\mu$. The induced monopole
currents are the straight lines with the direction $n_\mu$. Many--instanton
configurations induce abelian monopoles as well.

The calculations in the lattice gluodynamics show, that
the abelian monopoles are responsible for the confinement in the MA
projection \cite{Suz93}. The contribution of the abelian monopoles to the
Wilson loop leads to the area law with the correct string tension
coefficient \cite{StNeWe94,ShSu94}. Moreover, it was observed
\cite{EjKiMaSu95} that the long monopole loops alone are responsible for the
behavior of the string tension in the confinement phase of $SU(2)$ QCD. We
found that the monopole trajectories of this type may be induced by
instantons (and BPS monopoles) into MA projected $SU(2)$ vacuum. Therefore
the monopoles induced by instantons may in principle be important for the
confinement scenario.

\section*{Acknowledgments}

We have benefited from many useful discussions with D.I.~Diakonov and
M.I.~Polikarpov. We gratefully acknowledge M.I.Polikarpov for a critical
reading of the paper. This work was supported by the Grant No. MJM300,
financed by the International Science Foundation and the Government of the
Russian Federation, the JSPS Program on Japan -- FSU scientists
collaboration, the Grant INTAS-94-0840 and by the Grant No.  93-02-03609,
financed by the Russian Foundation for Fundamental Sciences.

\end{document}